\documentclass[11pt,a4paper]{article}
\usepackage{authblk}

\usepackage{graphicx}     
\usepackage{enumerate}
\usepackage{natbib}       
\usepackage{amsmath}
\usepackage{amsthm}            
\usepackage{amssymb}           
\usepackage{tabularx}          
\usepackage{enumitem}          
\usepackage[section]{placeins} 

\theoremstyle{theorem}\newtheorem{definition}{Definition}
\theoremstyle{remark}

\title{A Non-equilibrium Approach to Model Flash Dynamics with Interface Transport} 

\author[1]{Aar\'on Romo-Hernandez} 
\author[2]{Nicolas Hudon}
\author[3]{B. Erik Ydstie}
\author[1]{Denis Dochain}

\affil[1]{ICTEAM, Universit\'e Catholique de Louvain, B-1348, Louvain-la-Neuve, Belgium (e-mail: aaron.romo,denis.dochain@uclouvain.be).}
\affil[2]{Department of Chemical Engineering, Queen's University, Kingston, ON, K7L 3N6 Canada (e-mail: nicolas.hudon@queensu.ca).}
\affil[3]{Department of Chemical Engineering, Carnegie Mellon University, Pittsburgh, PA 15213, USA (e-mail: ydstie@cmu.edu).}

\begin{document}
\maketitle

\begin{abstract}       
  \noindent \rule{0.85\textwidth}{0.75pt}
  
  \vspace{3pt}
  \noindent This article presents a modeling framework for a class of multiphase chemical systems based on non-equilibrium thermodynamics.  Compartmental modeling is used to establish the dynamic properties of liquid-vapor systems operating far from thermodynamic equilibrium. In addition to the bulk-phase molar/energetic dynamics, interface transport processes yield to algebraic constraints in the model description.  The irreversible system is thus written as a system of Differential-Algebraic Equations (DAEs).  The non-equilibrium liquid-vapor DAE system is proven to be of index one.  A local stability analysis for the model shows that the equilibrium state is unstable for non-isobaric operation regimes, whereas numerical evidence shows that isobaric operation regimes are stable.  To extend the stability analysis, internal entropy production for the irreversible flash-drum is presented as a Lyapunov function candidate.

  \vspace{10pt}
  
  
  \noindent \rule{0.85\textwidth}{0.75pt}
\end{abstract}

\newpage
\section{Introduction}

Despite its historical and contemporary significance, design, control and operation of multiphase processes have been challenging tasks in process systems engineering over the years \citep{skogestad1997distillation,taylor2000modelling}.  In this article, we explore the possibility of modeling a special class of multiphase systems, the open flash-drum, utilizing a non-equilibrium physics-based perspective.  In contrast with traditional equilibrium formulations, non-equilibrium models can keep track of irreversible phenomena such as energy degradation and entropy production \citep{Mazur1984}.  Moreover, irreversible schemes have led to useful insights for stability analysis and feedback control design in the chemical process systems literature \citep{alonso1996process, ydstie1997process,favache2009thermodynamics,garcia_sandoval2015,ydstide2016stability}.  The work presented in this article expands the available modeling techniques and stability theory to include open liquid-vapor systems that operate far from thermodynamic equilibrium. 

Dissipative systems theory, originally proposed by \citet{willems1972dissipative} as an extension of (linear) passivity-based analysis, has been established as a fundamental tool for analysis and control design of mechanical and electrical systems.  As for chemical systems, dissipative systems theory has received an increasing level of attention after Alonso and Ydstie \citep{alonso1996process,ydstie1997process} developed the concept in detail using the first and second laws of thermodynamics.  A decade later,  in the article by \citet{favache2009thermodynamics}, the possibility of characterizing the continuous stirred tank reactor (CSTR) through energetic and entropic formulations is explored with insightful results. Other worth mentioning references include the work by \citet{HOANG2012}, where Lyapunov control laws, based on thermodynamic availability, are proposed for a larger class of non equilibrium CSTRs; and the article by \citet{garcia_sandoval2015}, where dissipative properties, based on internal entropy production, are established for irreversible CSTRs.  

Thermodynamic process systems theory has focused mostly on single-phase systems, also known as simple thermodynamic systems.  And it is not until recently that multiphase systems theory analysis has gained an increasing interest from the scientific community.  
\begin{figure}[h!]
  \centering
  \includegraphics[width=0.45\linewidth]{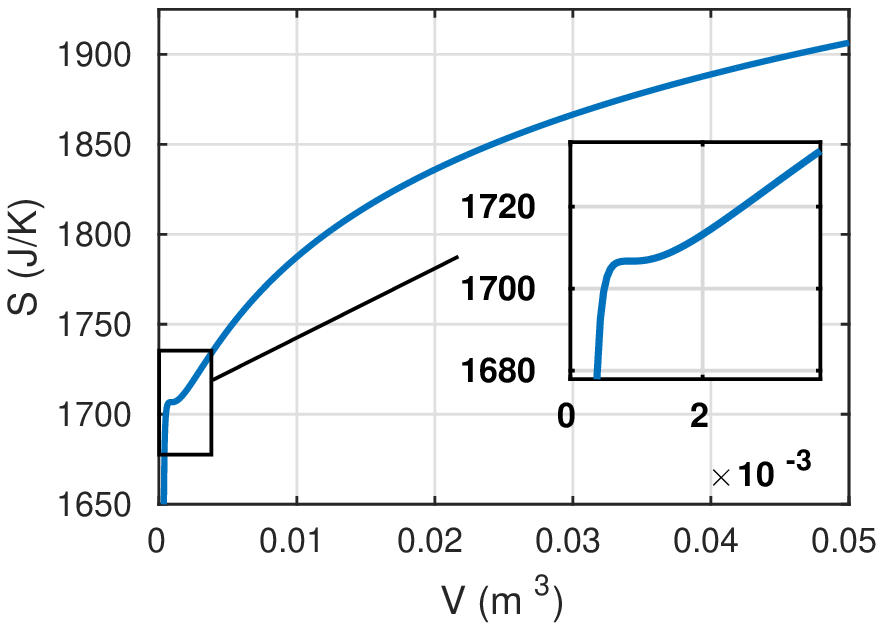}
  \includegraphics[width=0.45\linewidth]{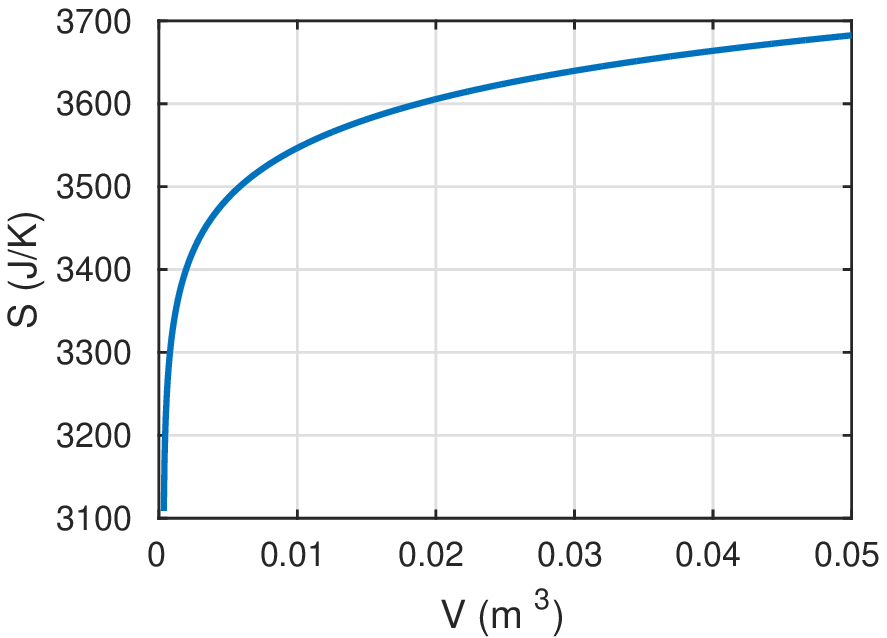}
  \caption{Van der Waals entropy for water: liquid-vapor (left) and gas (right).}
  \label{fig:entropy}
\end{figure}
For simple thermodynamic systems, entropy (resp., internal energy) is a strictly concave (resp., convex) function of the extensive variables \citep{C1985thermodynamics}, as depicted Figure~\ref{fig:entropy} (right) below.  Such concavity properties had permitted to assess for stability using Lyapunov theory and dissipative analysis \citep{favache2009thermodynamics,garcia_sandoval2015,HOANG2012}.  For multiphase systems the concavity is not strict \citep{gromov2012hybrid} as the very existence of separate phases follows as a consequence of the loss of concavity in entropy \citep{C1985thermodynamics}, see Figure~\ref{fig:entropy} (left) below.  To characterize a system regardless of its convexity properties, we propose to study liquid-vapor processes based on internal entropy production rather than entropy or internal energy potentials.

Historically, systems theory analysis for multiphase processes can be traced back to the pioneering work of \citet{rosenbrock1963lyapunov}, who demonstrated that a non-ideal binary distillation column operates at a unique asymptotically stable steady state. Three decades later \citet{rouchon1993geometry} developed a stability analysis, based on geometric considerations, for a multicomponent flash-drum.  Their approach unfortunately does not extend to multistage process units. In a more recent contribution, \citet{ydstide2016stability} developed conditions for the existence of a unique stable steady-state for an adiabatic flash-drum operating on an equilibrium manifold.  Looking forward to extend the process systems theory to include dynamic non-equilibrium liquid-vapor units, a modeling framework is presented in our previous contribution \cite{romohernandez2018nonequilibrium}.  The article discussed here serves as an extension to \cite{romohernandez2018nonequilibrium} that includes the construction of the entropy production function for the irreversible flash-drum.  As a Lyapunov function candidate, entropy production can serve to characterize the dynamic properties for nonlinear irreversible flash-drums.  In addition, details on mathematical procedures excluded from \cite{romohernandez2018nonequilibrium} are included thorough different sections in the work presented below.  

In this article, we consider a non-equilibrium flash-drum as the combination of three subsystems: a liquid phase, a gas phase, and an interface. To study such system, the  paper is organized as follows.  In Section \ref{sec:liqu-vapor-thermo-systems}, the non-equilibrium liquid-vapor model is built as a system of differential-algebraic equations (DAEs) from mass and energy conservation principles. The DAE model takes into account exchange processes through the interface, viewed as constraints on the dynamics of the system.  In Section \ref{sec:entropy-production-1}, internal entropy production for the flash-drum is written as the sum of products of flows and driving forces.  Being positive definite, internal entropy production is a physics-based Lyapunov function candidate to characterize the stability of the irreversible flash-drum.  Numerical results are presented in Section \ref{sec:numerical-simulation}.  First, the DAE model is rewritten and improved for numerical integration using a bijective change of coordinates.  Second, a non-ideal water-methanol liquid gas mixture is considered to present dynamic trajectories and results regarding Lyapunov's first and second methods for stability.  Conclusions and future areas for research are discussed in Section \ref{sec:concl-future-work}.

\section{Liquid-Vapor Thermodynamic Systems}\label{sec:liqu-vapor-thermo-systems}

In this section, we develop a system of nonlinear differential-algebraic equations (DAEs) of index one that describes a flash-drum as the interconnection of two thermodynamic subsystems.  We motivate the need for such description through Gibbs equation as a way to describe entropy variations inside an open liquid-vapor system. The obtained DAE model is an abstract representation of the first law of thermodynamics applied to the irreversible flash-drum.

A thermodynamic system is completely defined once its physical properties are determined. For instance, all the physical properties of a closed chemical system with $c$ chemical components can be recovered once a parti\-cular set of coordinates  $(U,V,N_1,\ldots,N_c)$ is known.  Such coordinates represent the internal energy, volume, and mole numbers respectively.  This is known as the first postulate of thermodynamics and it was proposed by \citet{C1985thermodynamics}, on the basis of the Gibbs' school on modern thermodynamics.


The second postulate of thermodynamics defines entropy as a function that is maximized at thermodynamic equilibrium \citep{C1985thermodynamics}.  Moreover, such function relates the entropy with the intensive variables \citep{wightman1979convexity}.  

\begin{definition}[Entropy]
Let $(U,V,N_1,\ldots,N_j)\in\mathbb R_{> 0}^{c+2}$ represent the internal energy, volume, and mole numbers for a closed thermodynamic system, and let $(T,P,\mu_1,\ldots,\mu_c)$ stand as the temperature, pressure and the chemical potential.  Then, entropy is a concave function
\begin{subequations}\label{eq:formal_structure}
    \begin{align}
      S = \text S (U,V,N_1,\ldots,N_c)\label{eq:entropy}
    \end{align}
  that is maximized over the equilibrium states of the thermodynamic system. The function $\text S(\cdot)$ is said to be homogeneous of degree one, \textit{i.e.},
    \begin{align}    
       \text S (\lambda U, \lambda V, \lambda N_1,\ldots, \lambda N_c) =  \lambda \text S (U,V,N_1,\ldots,N_c), \phantom{...} \forall \lambda > 0. \nonumber
    \end{align}
  Entropy is at least once differentiable.  The derivatives of $\text S(\cdot)$ satisfy
    \begin{align}
      \dfrac{\partial S}{\partial U}   = \dfrac{1}{T}, \;\;
      \dfrac{\partial S}{\partial V}   = \dfrac{P}{T}, \;\;
      \dfrac{\partial S}{\partial N_j} = \dfrac{-\mu_j}{T}, \;\; j\in\{1,\ldots,c\}. \label{eq:conjugates}
    \end{align}
\end{subequations}
As a consequence of \eqref{eq:conjugates}, the coordinates $(T, P,\mu_1, \ldots, \mu_c)$ are said to be the conjugates to the extensive variables  $(U,V,N_1, \ldots, N_c)$.
\mbox{}\hfill $\Box$
\end{definition} 

The formal structure of thermodynamics, Equation \eqref{eq:formal_structure}, is defined for closed systems at thermodynamic equilibrium \citep{C1985thermodynamics}.  Nevertheless, Equation \eqref{eq:formal_structure} is still considered valid locally when studying systems that are not at thermodynamic equilibrium \citep{HOANG2012}.  Non-homogeneous systems, non-stationary processes and open vessels exchanging mass and energy with the environment are some examples of non-equilibrium systems.  Computing the differential of \eqref{eq:entropy} we obtain what is known as the Gibbs equation 
  \begin{align}
    \text dS = \frac{1}{T} dU + \frac{P}{T} dV + \frac{\mu_1}{T} dN_1 + \ldots + \frac{\mu_c}{T} dN_c. \label{eq:entropy_differential}
  \end{align}
When time variations are considered instead of the differentials in \eqref{eq:entropy_differential} we get 
  \begin{align}
    \dfrac{dS}{dt} = \frac{1}{T} \dfrac{dU}{dt} + \frac{P}{T} \dfrac{dV}{dt} + \frac{\mu_1}{T} \dfrac{dN_1}{dt} + \ldots + \frac{\mu_c}{T} \dfrac{dN_c}{dt}. \label{eq:Gibbs_open}
  \end{align}
As the derivatives on the right hand side of~\eqref{eq:Gibbs_open} can represent exchange rates between a system and its environment, Equation~\eqref{eq:Gibbs_open} is frequently considered as an extension of Gibbs equation \eqref{eq:entropy_differential} for open thermodynamic systems.  In the following section, we discuss how to take advantage of conservation laws to describe the exchange rates for $(U,V,N_1,\ldots,N_c)$ that appear on the right hand side of~\eqref{eq:Gibbs_open}, particularly for the open flash-drum. 

\subsection{Conservation Principles}
\label{sec:cons-princ}

\begin{figure}[h]
  \centering
  \includegraphics[width=0.45\linewidth]{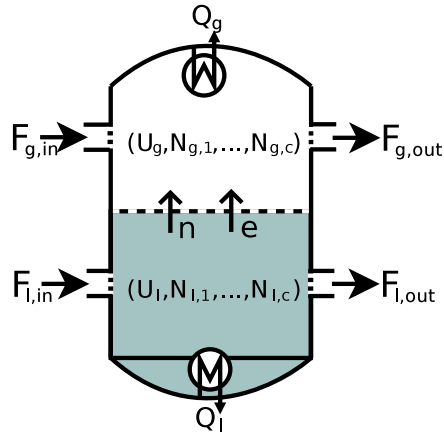}
  \caption{Liquid-vapor open thermodynamic system with interface transport}
  \label{fig:open_thermo_system}
\end{figure}

From here on, we consider the system of interest as a flash-drum with one liquid phase on the bottom of a rigid vessel and a vapor phase above as depicted in Figure~\ref{fig:open_thermo_system}.  Inside the system, there are $c$ chemical components distributed between phases.  To avoid unnecessary repetition, we set sub-index $\alpha \in \{l,g\}$ as the phase sub-index. 

Each phase inside the flash-drum has certain amount of internal energy $U_\alpha$, and a determined number of moles $N_{\alpha,j}$ for each component $j\in \{1,\ldots,c\}$.  As the system is open, moles flow in and out at convective rates $F_{\alpha, N_j, \mathrm{in}}$ (mol/sec) and $F_{\alpha, N_j, \mathrm{out}}$ (mol/sec) respectively.  Thermal energy is exchanged between phase $\alpha$ and the environment at a rate $Q_\alpha$ (J/sec) through a heat exchanger.  Additionally, moles and energy can be exchanged between phases.  We set the interface energy exchange rate as $e$ (J/sec) and the interface molar exchange rate as $n$ (mol/sec).  These represent the rates at which total energy and total mole numbers are exchanged from the liquid to the gas phase.  As a final consideration, we set $K_\alpha$ to be the kinetic energy of the flow moving through phase $\alpha$.

For modeling purposes, the following assumptions are considered:%
\begin{enumerate}[itemsep=1pt,label=A\arabic*.]
\item Each phase is perfectly mixed.\label{item:1}
\item Potential energy is constant all over the process.\label{item:2}
\item Flow compressibility and viscous losses are negligible.
\item Kinetic energy variations are insignificant when compared to variations in enthalpy or in internal energy.\label{item:4}
\item No moles or energy accumulate in the interface.\label{item:5}
\item The liquid phase, the vapor phase, and the interface operate locally at a state of  thermodynamic equilibrium.\label{item:6} 
\end{enumerate}
It follows from Assumption \ref{item:1} that molar dynamics for component $j\in\{1,...,c\}$ correspond to%
\begin{subequations}\label{eq:molar_balance}
    \begin{align}
      \dfrac{d N_{g,j}}{dt}    
      &= F_{g,N_j,\mathrm{in}} - \frac{N_{g,j}}{V_g} F_{g,V,\mathrm{out}} + n_{g,j}\label{eq:molar_balance_gas}\\
      \dfrac{d N_{l,j}}{dt}   
      &= F_{l,N_j,\mathrm{in}} - \frac{N_{l,j}}{V_l} F_{l,V,\mathrm{out}} - n_{l,j}.\label{eq:molar_balance_liquid}
    \end{align}
\end{subequations}
Under Assumptions \ref{item:2}-\ref{item:4}, the internal energy can be written using a total energy balance%
\begin{subequations}\label{eq:internal_energy_balance}
    \begin{align}
      \dfrac{d U_g}{dt}   
      &= F_{g,H,\mathrm{in}} - \frac{U_g}{V_g} F_{g,V,\mathrm{out}} - P_{g} F_{g,V,\mathrm{out}} - P_g \dot V_g + Q_g + e_{g} \label{eq:internal_energy_balance_gas}\\
      \dfrac{d U_l}{dt}   
      &= F_{l,H,\mathrm{in}} - \frac{U_l}{V_l} F_{l,V,\mathrm{out}} - P_{l} F_{l,V,\mathrm{out}} - P_l \dot V_l + Q_l - e_{l}, \label{eq:internal_energy_balance_liquid}
    \end{align}
\end{subequations}
and, kinetic energy can be described using a balance on mechanical energy (derived from a momentum balance)%
\begin{subequations}\label{eq:mechanical_energy_balance}
    \begin{align}
      \dfrac{d K_g}{dt}  
      &= F_{g, K, \mathrm{in}} - \frac{K_g}{V_g} F_{g,V,\mathrm{out}} - P_g \dot V - P_g F_{g,V,\mathrm{out}} + e_{g,K}\\
      \dfrac{d K_l}{dt}  
      &= F_{l, K, \mathrm{in}} - \frac{K_l}{V_l} F_{l,V,\mathrm{out}} - P_l \dot V - P_l F_{l,V,\mathrm{out}} - e_{l,K}.
    \end{align}
\end{subequations}
In equations \eqref{eq:molar_balance}-\eqref{eq:mechanical_energy_balance}, $n_{\alpha, j}$, $e_\alpha$ and $e_{\alpha, K}$ represent interface sources/sinks for moles (in component $j$) and energy (total and mechanical).  Convective inflow rates in \eqref{eq:molar_balance}-\eqref{eq:mechanical_energy_balance}, $F_{\alpha, N_j, \mathrm{in}}$, $F_{\alpha, H, \mathrm{in}}$, $F_{\alpha, V, \mathrm{in}}$, $F_{\alpha, K, \mathrm{in}}$, are assumed as fixed inputs.  As there is not a conservation principle for volume \citep{T1993masstransfer}, we write the liquid volume as a function of the molar holdup \citep{sandler}
\begin{subequations}\label{eq:volume}
    \begin{align}
      V_l = \bar {\text v}_1 N_{l,1} + \cdots + \bar {\text v}_c N_{l,c},\label{eq:volume_liquid}
    \end{align}
  where $\bar {\text v}_j$ corresponds to the partial molar volume of
  component $j$ in the liquid phase.  As the vessel containing both
  phases is rigid, the gas volume corresponds to the volume not occupied
  by the liquid phase
    \begin{align}
      V_g = V_{o} - V_l,\label{eq:volume_gas}
    \end{align}
\end{subequations}
where $V_{o}$ is a constant representing the volume of the rigid unit.  Volumetric outflow rates are described as a function of the phase flow velocity $v_\alpha(\cdot)$, which in turn depends on the kinetic energy of the phase%
  \begin{align}
    F_{\alpha,V,\mathrm{out}} &= A_{\alpha,\mathrm{out}} \;v_\alpha(K_\alpha,N_\alpha), \phantom{.......} v_\alpha = \sqrt{2 K_\alpha/M_\alpha(N_{\alpha})}, \label{eq:kinetic_energy}
  \end{align}
where $A_{\alpha,\mathrm{out}}$ stands as a parameter that represents the cross sectional area of the flow line, and $M_\alpha = \sum \bar m_j N_{\alpha,j}$ represents the mass holdup in bulk-phase $\alpha$.  Heat flows $Q_\alpha$ in the energy balance equation \eqref{eq:internal_energy_balance} are written as being proportional to differences between the heat exchanger temperature $T_{\alpha,Q}$  and the bulk-phase temperature $T_\alpha$ 
  \begin{align}
    Q_\alpha = \lambda_\alpha (T_{\alpha,Q} - T_\alpha), \label{eq:heat_exchanger}
  \end{align}
where $\lambda_\alpha$ is a known heat exchange parameter.
Temperature in turn is related to the internal energy.  The internal energy for 
phase $\alpha$ can be written as%
  \begin{align}
    U_\alpha  = U_{o,\alpha}(N_\alpha) + \mathcal C_{\alpha}(N_\alpha) (T_\alpha - T_{\text o}),\label{eq:internalEnergy}
  \end{align}
%
where
$U_{o,\alpha}(\cdot) = \sum \bar{u}_{\text o, j} N_{\alpha,j}$ represents the internal energy of the system at a reference temperature $T_{\text o}$, and $\mathcal C_{\alpha}(\cdot) = \sum N_{\alpha,j} \, \bar c_{\alpha,j}$ stands for the total heat capacity for phase $\alpha$.  The ideal gas equation is used to write the pressure in the gas phase as a function of the extensive parameters 
\begin{subequations}\label{eq:P}
    \begin{eqnarray}
      P_{\text g}(U_g,V_g,N_g)
      &= & R \dfrac{N_g}{V_g} \bigg( T_{\text o} + \frac{ U_g - U_{o,g}(N_g) }{\mathcal C_g(N_g)} \bigg).\label{eq:P_g}
    \end{eqnarray}
  An accurate description for the liquid pressure would require a detailed study on the hydrodynamic properties of the liquid subsystem, which is beyond the scope of this article.  The interested reader is referred to the work of \citet{teixeira2017}.  As the system is contained in a rigid vessel we assume, from Pascal's principle, that the interface, the liquid, and the gas phases are at the same (not necessarily constant) pressure.
    \begin{align}
      P_l = P_i = P_g(U_g,V_g,N_g).\label{eq:P_l}
    \end{align}
\end{subequations}

Equations \eqref{eq:molar_balance}-\eqref{eq:mechanical_energy_balance} represent conservation principles \citep{BSL2002transport}.  The conservation principles put together with its constitutive equations \eqref{eq:volume}-\eqref{eq:P} stand as a standard representation of an open thermodynamic system.  It should be noted here that the description is not yet completed as the interface exchange rates are not included in the constitutive equations \eqref{eq:volume}-\eqref{eq:P}.  To complete the flash-drum system description, in the next section we describe molar $n_{\alpha, j}$, energetic $e_\alpha$ and mechanical $e_{\alpha, K}$ interface exchange rates as function of the intensive variables.

\subsection{Interface Exchange Rates}\label{sec:interf-exch-rates}

The model presented so far considers the liquid and the gas phases as separated subsystems inside the multiphase unit.  As a consequence, the flash-drum is not required to evolve over an equilibrium manifold and inhomogeneities in temperature and chemical potential can arise between phases.  The ideas behind the modeling for non-homogeneous systems were first presented by \citet{Krishnamurthy1985nonequilibrium}.  Their model relies on describing interfacial temperature, pressure, and compositions solely based on Assumptions \ref{item:5} and \ref{item:6} plus a mechanical equilibrium assumption.  The non-equilibrium model of \citet{Krishnamurthy1985nonequilibrium} is unfortunately limited to stationary regimes.  Below, we reproduce such description adapted to our open process.

\begin{figure}[t!]
  \centering
  \includegraphics[width=0.75\linewidth]{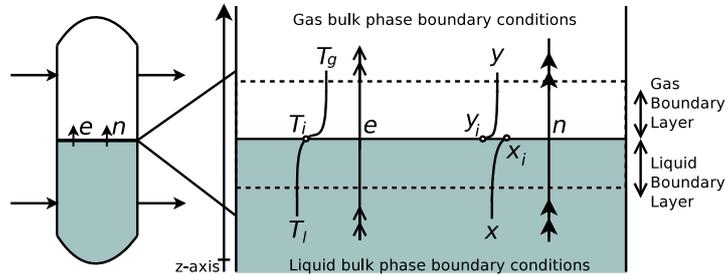}
  \caption{Interface inside a non-equilibrium liquid-vapor System}
  \label{fig:interface}
\end{figure}
Two boundary layers are assumed to surround the gas-liquid interface as depicted in Figure~\ref{fig:interface}.  Liquid and gas coexist in the interface between the layers at local equilibrium.  Subindex $i$ is set to describe local interfacial temperature, and molar fractions. In addition, $n$ is defined as the interface molar exchange which represents the total rate at which moles flow from liquid to gas phase.  The interface variables are therefore referred as the vector
  \begin{align}
    [T_i,\;  x_{1,i},\; y_{1,i}, \ldots, x_{c,i},\; y_{c,i}, n]^{\text t}.\label{eq:interface_variables}
  \end{align}

Temperature and composition gradients are allowed between the bulk-phases and the interface, see Figure~\ref{fig:interface}. These gradients are known to be the driving forces behind interface flow rates.  In the molar balance equation \eqref{eq:molar_balance}, the rate at which chemical component $j$ flows from phase $\alpha$ towards/from the interface is represented by $n_{\alpha,j}$.  Neglecting the effects of temperature, pressure gradients, and intercomponent diffusive transport, we can write \citep{T1993masstransfer}
\begin{subequations}\label{eq:interface_molar_exchange}
    \begin{align}
      n_{g,j} &=  k_{g,j} \, C_g (y_{j,i} - y_j) + n_{g} \, y_j \\
      n_{l,j} &=  k_{l,j} \, C_l (x_j - x_{j,i}) + n_{l} \, x_j, 
    \end{align}
\end{subequations}
where $y_j, \; x_j$ and $C_\alpha$ represent respectively the molar fractions and concentrations in the bulk-phases.  The parameter $k_{\alpha, j}$ is assumed to be known and stands for a diffusive transport parameter.  The total interface transport rate that appears in  \eqref{eq:interface_molar_exchange} corresponds simply to
  \begin{align}
    n_\alpha = \sum_{j=1}^c n_{\alpha,j}. \nonumber
  \end{align}
The rate at which energy flows from phase $\alpha$ towards/from the interface in Equation \eqref{eq:internal_energy_balance} is represented by $e_\alpha$.  The interface energy flow rate can be written as the sum of convective and thermal energy contributions \citep{T1993masstransfer,BSL2002transport}
\begin{subequations}\label{eq:interface_energy_exchange}
    \begin{align}
      e_g
      &= \sum_{j=1}^c n_{g,j} \; \bar h_{g,j} + \lambda_{g,i} (T_{i} - T_{g}) \\
      e_l    
      &= \sum_{j=1}^c n_{l,j} \; \bar h_{l,j} + \lambda_{l,i} (T_{l} - T_{i}), 
    \end{align}
\end{subequations}
where $\bar h_{\alpha,j}$ stands as the partial molar enthalpy of component $j$ in phase $\alpha$, $\lambda_{\alpha, i}$ is a known thermal exchange parameter, $T_\alpha$ represents the bulk-phase temperature and $T_i$ the interfacial temperature.  The rate at which mechanical energy is exchanged between bulk-phase and the interface in the mechanical energy equation \eqref{eq:mechanical_energy_balance} is represented as $e_{\alpha, K}$. This term  corresponds to \citep{BSL2002transport}
  \begin{align}\label{eq:interface_kinetic_exchange}
    e_{\alpha, K} 
    &= \frac{1}{2} v_{\,\alpha,\text i}^2 m_{\alpha} + \frac{P_{\,\alpha}}{\rho_{\,\alpha}} m_{\alpha},
  \end{align}
where $\rho_{\,\alpha}$ stands as the density of the bulk-phase $\alpha$. Setting $\bar m_j$ to be the molar mass of component $j$, we write $m_{\alpha} = \sum \bar m_j \, n_{\alpha,j}$ for the mass interface transport rate.  The term $v_{\alpha, i}$ represents the average velocity of the mass flowing through the boundary layers surrounding the interface.  Setting $\text A_i$ as the interface area we define the interface flow velocity in boundary layer $\alpha$ as
  \begin{align}
    v_{\,\text i,\alpha} = \dfrac{m_{\alpha}}{\rho_\alpha \text A_i}.\nonumber
  \end{align}

It should be noted that the interface exchange rates \eqref{eq:interface_molar_exchange}-\eqref{eq:interface_kinetic_exchange} depend on the $2c + 2$ interface variables written in \eqref{eq:interface_variables}.  These interface variables are recovered as the solution to an algebraic system of equations built from Assumptions \ref{item:5} and \ref{item:6}  

\subsection{Interface Algebraic System}
\label{sec:interf-algebr-syst}

No accumulation of moles in the interface, Assumption \ref{item:5}, gives $c-1$ equations%
  \begin{align}\label{eq:interface_mass_conservation}
    k_{g,j} \, C_g (y_{j,i} - y_j) -  k_{l,j} \, C_l (x_j - x_{j,i}) + (y_j -x_j)n = 0, \;\; &j=1,\ldots,c-1,
  \end{align}
where $n := n_l = n_g$ is the total molar interface rate defined in~\eqref{eq:interface_variables}.  No accumulation of energy in the interface, Assumption \ref{item:5}, leads to one more algebraic restriction%
  \begin{align}\label{eq:interface_energy_conservation}
    \sum_{j=1}^c n_{j} \Delta \bar h_{\mathrm{vap},j} + \lambda_{g,i} (T_{g} - T_{i}) - \lambda_{l,i} (T_{l} - T_{i}) = 0, 
  \end{align}
where $n_j := n_{l,j} = n_{g,j}$ represents the molar interface rate for component $j$ and  $\Delta \bar h_{\mathrm{vap},j} = \bar h_{g,j} - \bar h_{l,j}$ represents the partial enthalpy of vaporization for component $j$.  To complete the interface description we add $c+2$ interface equilibrium equations%
  \begin{subequations}\label{eq:interface_equilibrum}
    \begin{align}
      0 &= y_{j,i} - K_j(T_i,x_{1,i},\ldots,x_{c,i}) x_{j,i} \phantom{....} &j=1,\ldots,c \\
      0 &= 1 - \textstyle\sum_{j=1}^c x_{j,i} \\
      0 &= 1 - \textstyle\sum_{j=1}^c y_{j,i}, 
    \end{align}
  \end{subequations}
where $K_j(\cdot)$ represents the liquid-vapor composition ratio for component $j$.  Even though the liquid-vapor composition ratio is frequently assumed constant, this term is, in general, a nonlinear function of the intensive interface variables, see Equation~\eqref{eq:twoComponentEquilibrium} in~\ref{sec:example_margules}.

Balance equations \eqref{eq:molar_balance}-\eqref{eq:mechanical_energy_balance} restricted by the interface equations \eqref{eq:interface_mass_conservation}-\eqref{eq:interface_equilibrum} sum up to $4c + 6$ nonlinear coupled differential algebraic equations that represent a semi-explicit DAE system%
  \begin{subequations}\label{eq:DAE}
    \begin{align}
      \dot{\bf z} &= f( {\bf z}, {\bf w}) \label{eq:differential}\\
      0 &= g({\bf z}, {\bf w}), \label{eq:algebraic}
    \end{align}
  \end{subequations}
where ${\bf z}$ refers to bulk-phase variables
\begin{subequations}
    \begin{align}
      {\bf z} = [ N_{g,1} \ldots N_{g,c} \;\; N_{1,l} \ldots N_{l,c} \;\; U_g \;\; U_l \;\; K_{g} \;\; K_{l} ]^{\,\text t}, 
    \end{align}
  and ${\bf w}$ represents the interface variables. %
    \begin{equation}
      {\bf w} = [ y_{1,i} \ldots y_{c,i} \;\; x_{1,i} \ldots x_{c,i} \;\; T_i \;\; n ]^{\,\text t}. 
    \end{equation}
\end{subequations}

The index of a DAE system is the number of times we have to differentiate the constrains $g(\cdot)$ to put aside the algebraic restrictions and write the DAE as an equivalent ordinary differential equations (ODEs) system.  A DAE system has index one if and only if the Jacobian $J_{\bf w} (g)$ has full rank \citep{Brenan_1996ch2}.  It follows from the implicit function theorem that, as $J_{\bf w} (g)$ is non-singular, there exists a unique function $h({\bf z})$ such that substitution of ${\bf w} = h({\bf z})$ satisfies the algebraic restriction in the DAE, \textit{i.e.}, $g({\bf z}, h({\bf z}) ) = 0$.  Therefore, if a DAE system is of index one we can virtually solve the algebraic part through the mapping $ {\bf w} = h({\bf z})$.  Substitution of ${\bf w} = h({\bf z})$ into the differential part would transform a DAE model to an equivalent ODE system.

The Jacobian $J_{\bf w} (g)$ for equation \eqref{eq:algebraic} is non singular (see \ref{sec:jacobian-g}), therefore system \eqref{eq:DAE} is of index one.  The irreversible flash-drum DAE model is thus equivalent to an ODE system and it can be characterized using Lyapunov theory.   In the next section we put together Gibbs equation \eqref{eq:Gibbs_open} with the balance equations in \eqref{eq:differential} to compute variations in entropy with respect to time.  This forms the basis to calculate the entropy production, a positive definite thermodynamic potential, considered as a Lyapunov function candidate used to characterize the irreversible flash-drum.

\section{Entropy Production}
\label{sec:entropy-production-1}

In this section, we compute the internal entropy production for the irreversible flash-drum using an entropy balance equation.  Being positive definite, the internal entropy production represents a physics-based Lyapunov function candidate which we can use to characterize the dynamics for the nonlinear flash-drum system \eqref{eq:DAE}.

Entropy is an extensive non-conserved property \citep{sandler}.  The entropy production rate $\sigma$ for the open flash-drum can be written from an entropy balance as 
  \begin{align}
    \sigma = \sum_{\alpha\in\{g,l\}} \bigg( F_{\alpha, S, \mathrm{out}} - F_{\alpha, S, \mathrm{in}} - \frac{Q_{\alpha}}{T_{\alpha, Q}} \bigg) + \frac{dS}{dt}, \label{eq:entropy_production}
  \end{align}
where $F_{\alpha, S, \mathrm{in}}$ and $F_{\alpha, S, \mathrm{out}}$ represent convective flow rates of entropy, and $Q_{\alpha}$ stands as a heat source at temperature $T_{\alpha, Q}$.  Because entropy is additive over subsystems \citep{C1985thermodynamics}, entropy variations with respect to time can be written using equation \eqref{eq:Gibbs_open} as
  \begin{align}
    \frac{dS}{dt} = \sum_{\alpha\in\{g,l\}} \frac{dS_\alpha}{dt} = \sum_{\alpha\in\{g,l\}} \frac{1}{T_\alpha} \dfrac{dU_\alpha}{dt} + \frac{P_\alpha}{T_\alpha} \dfrac{dV_\alpha}{dt} + \frac{ - \boldsymbol \mu_{\alpha}^{\text t}}{T_{\alpha}} \dfrac{d{\bf N}_\alpha}{dt},\label{eq:Gibbs_liquid_vapor}
  \end{align}
where the vector notation
  \begin{align}
    \boldsymbol \mu_\alpha = [\mu_{\alpha, 1} \; \ldots \; \mu_{\alpha, c}]^{\text t}, \phantom{.....}
    {\bf N}_\alpha = [N_{\alpha, 1} \; \ldots \; N_{\alpha, c}]^{\text t}, \nonumber
  \end{align}
has been introduced.  Note that $\dot {\bf N}_\alpha$ and $\dot U_\alpha$ in  \eqref{eq:Gibbs_liquid_vapor} represent the molar and energy balances for the flash-drum, equations \eqref{eq:molar_balance}--\eqref{eq:internal_energy_balance}, and $\dot V_\alpha$ can be recovered from Equation \eqref{eq:volume}.  

Entropy for each phase is a homogeneous of degree one function, Equation~\eqref{eq:formal_structure}.  Thus, integration of $dS_\alpha$, see Equation~\eqref{eq:entropy_differential}, gives \citep[Euler's theorem]{C1985thermodynamics} 
  \begin{align}
    S_\alpha = \frac{1}{T_\alpha} U_\alpha + \frac{P_\alpha}{T_\alpha} V_\alpha + \dfrac{\boldsymbol \mu_\alpha^{\text t}}{T_\alpha} {\bf N}_\alpha, \;\;\;\;\;\; \alpha\in\{g,l\}. \label{eq:entropy_euler_alpha} 
  \end{align}
Each bulk-phase is perfectly mixed, Assumption~\ref{item:1} Then, multiplying \eqref{eq:entropy_euler_alpha} by the inverse of the residence time $1/\tau_\alpha = F_{\alpha, V, \mathrm{out}} / V_{\alpha}$ we can write the outflow rate of entropy as
\begin{subequations}\label{eq:entropy_flows}
    \begin{align}
      F_{\alpha,S,\mathrm{out}} = 
      \frac{1}{T_\alpha} F_{\alpha,U,\mathrm{out}} + \frac{P_\alpha}{T_\alpha} F_{\alpha, V, \mathrm{out}} + \frac{- \boldsymbol \mu_{\alpha}^{\text t}}{T_{\alpha}} {\bf F}_{\alpha,N,\mathrm{out}},  \label{eq:entropy_euler_alpha_outflows}
    \end{align}
  where $F_{\alpha,Z,\mathrm{out}}$ holds for the convective flow rate of property $Z$, $F_{\alpha,Z,\mathrm{out}} = \tilde Z_\alpha F_{\alpha, V, \mathrm{out}}, \, Z\in\{S, U, {\bf N}\}$, and $\tilde Z = Z/V$ represents a variable per unit of volume.  Rewritting equation \eqref{eq:entropy_euler_alpha_outflows} using inflow instead of outflow properties we can write the convective entropy inflow rates as
    \begin{align}
      F_{\alpha,S,\mathrm{in}} = 
      \frac{1}{T_{\alpha, \mathrm{in}}} F_{\alpha,U,\mathrm{in}} + \frac{P_{\alpha, \mathrm{in}}}{T_{\alpha, \mathrm{in}}} F_{\alpha, V, \mathrm{in}} + \frac{- \boldsymbol \mu_{\alpha, \mathrm{in}}^{\text t}}{T_{\alpha, \mathrm{in}}} {\bf F}_{\alpha,N,\mathrm{in}}.  \label{eq:entropy_euler_alpha_inflows}
    \end{align}
\end{subequations}
Substitution of \eqref{eq:Gibbs_liquid_vapor} and \eqref{eq:entropy_flows} into \eqref{eq:entropy_production} gives the entropy production for the irreversible flash-drum as%
  \begin{align}
    \sigma 
    &= \bigg( \dfrac{1}{T_g} - \dfrac{1}{T_{g,Q}} \bigg) Q_g + \bigg( \dfrac{1}{T_l} - \dfrac{1}{T_{l,Q}} \bigg) Q_l\nonumber \\
    & + \bigg( \frac{1}{T_{g}} - \frac{1}{T_{g,\mathrm{in}}} \bigg) F_{g,H,\mathrm{in}} + \bigg( \frac{1}{T_{l}} - \dfrac{1}{T_{l,\mathrm{in}}} \bigg) F_{l,H,\mathrm{in}} \nonumber \\
    & + \bigg( \frac{- \boldsymbol \mu^{\text t}_{g}}{T_{g}} - \frac{ - \boldsymbol \mu_{g,\mathrm{in}}^{\text t}}{T_{g,\mathrm{in}}} \bigg) {\bf F}_{g,N,\mathrm{in}} + \bigg( \frac{- \boldsymbol \mu^{\text t}_{l}}{T_{l}} - \dfrac{-\boldsymbol \mu^{\text t}_{l,\mathrm{in}}}{T_{l,\mathrm{in}}} \bigg) {\bf F}_{l,N,\mathrm{in}}. \nonumber \\
    & + \bigg( \frac{1}{T_{g}} - \frac{1}{T_{l}} \bigg) e + \bigg( \frac{- \boldsymbol \mu^{\text t}_{g}}{T_{g}} - \frac{- \boldsymbol \mu^{\text t}_{l}}{T_{l}} \bigg) {\bf n}. \label{eq:sigma}
  \end{align}
Each term in Equation \eqref{eq:sigma} can be identified as a source of entropy on a phenomenological basis:
\begin{itemize}
\item The first terms, 
    \begin{align}
      \bigg( \dfrac{1}{T_g} - \dfrac{1}{T_{g,Q}} \bigg) Q_g + \bigg( \dfrac{1}{T_l} - \dfrac{1}{T_{l,Q}} \bigg) Q_l, \nonumber
    \end{align}
  represent the entropy produced as heat $Q_\alpha$  is exchanged between phase $\alpha$ and the external heat sources.
\item The terms related with convective flow rates,
    \begin{align}
      \sum_{\alpha\in\{g,l\}} \bigg( \frac{1}{T_{\alpha}} - \frac{1}{T_{\alpha,\mathrm{in}}} \bigg) F_{\alpha,H,\mathrm{in}} + \bigg( \frac{- \boldsymbol \mu^{ t}_{\alpha}}{T_{\alpha}} - \frac{ - \boldsymbol \mu_{ \alpha,\mathrm{in}}^{ t}}{T_{ \alpha,\mathrm{in}}} \bigg) {\bf F}_{\alpha,N,\mathrm{in}}, \nonumber
    \end{align} 
  represent the entropy produced as the inflows are mixed with the bulk-phases.  Note that if inflows are at the same temperature and chemical potential (composition) as the bulk-phases, these terms are equal to zero.
\item The last two terms, 
    \begin{align}
      \bigg( \frac{1}{T_{g}} - \frac{1}{T_{l}} \bigg) e + \bigg( \frac{- \boldsymbol \mu^{\text t}_{g}}{T_{g}} - \frac{- \boldsymbol \mu^{\text t}_{l}}{T_{l}} \bigg) {\bf n}, \nonumber
    \end{align}
  represent the entropy produced as energy and moles are exchanged through the interface when phases are not homogeneous, \textit{i.e.}, far from thermodynamic equilibrium.
\end{itemize}
Note that when the flash-drum is completely isolated from the environment we have $F_{\alpha,H,\mathrm{in}} = 0$, ${\bf F}_{\alpha,N,\mathrm{in}} = 0$ and $Q_\alpha = 0$. Then, the entropy production is reduced to the internal entropy production 
  \begin{align}
    \sigma_i := \sigma|_{\mathrm{isolated}} = \bigg( \frac{1}{T_{g}} - \frac{1}{T_{l}} \bigg) e + \bigg( \frac{- \boldsymbol \mu^{\text t}_{g}}{T_{g}} - \frac{- \boldsymbol \mu^{\text t}_{l}}{T_{l}} \bigg) {\bf n}. \label{eq:sigma_i}
  \end{align}
Equation \eqref{eq:sigma_i} can be written as the sum of products between generalized flows and driving forces
  \begin{align}
    \sigma_i := \sigma|_{\mathrm{isolated}} = J_e \, e + J^{\text t}_{n,1} \, n_1 + \cdots + J^{\text t}_{n,c} \, n_c,\nonumber
  \end{align}
where the flows, see equations \eqref{eq:interface_molar_exchange}, \eqref{eq:interface_mass_conservation} and \eqref{eq:interface_energy_exchange},\eqref{eq:interface_energy_conservation}, correspond to%
  \begin{subequations}\label{eq:interface_flows}
    \begin{align}
      e   &:= \sum_{j=1}^c n_{g,j} \; \bar h_{g,j} + \lambda_{g,i} (T_{i} - T_{g}) = \sum_{j=1}^c n_{l,j} \; \bar h_{l,j} + \lambda_{l,i} (T_{l} - T_{i})  \label{eq:e_flow}\\
      n_j &:= k_{g,j} \, C_g (y_{j,i} - y_j) + n \, y_j =  k_{l,j} \, C_l (x_j - x_{j,i}) + n \, x_j,  \label{eq:nj_flow}
    \end{align}
  \end{subequations}
and the driving forces 
  \begin{align}
    J_e = \frac{1}{T_{g}} - \frac{1}{T_{l}}, \;\;\; {\bf J}^{\text t}_n = \frac{- \boldsymbol \mu^{\text t}_{g}}{T_{g}} - \frac{- \boldsymbol \mu^{\text t}_{l}}{T_{l}} \label{eq:dforces}
  \end{align}
come as a consequence of inhomogeneities between liquid and gas phases.  Even though the flows \eqref{eq:interface_flows} and the driving forces \eqref{eq:dforces} are not written using Onsager's relations \citep{prigogine1968introduction}, it can be easily verified that both \eqref{eq:interface_flows} and \eqref{eq:dforces} vanish at thermodynamic equilibrium.  Moreover, numerical evidence shows that $\sigma_i$ is positive definite and decreases with respect to time as the flash-drum reaches an equilibrium state in a system with fixed inflows, as shown in the next section.  

\section{Numerical Simulations}\label{sec:numerical-simulation}

In this section, a non-ideal methanol-water mixture is simulated to illustrate the properties of the model presented in Section \ref{sec:liqu-vapor-thermo-systems} and Section \ref{sec:entropy-production-1}.  First, a change of coordinates is used to transform the extensive-intensive description \eqref{eq:DAE} to an equivalent DAE system improved for numerical integration.  Then, a stability analysis based on Lyapunov first method is briefly discussed. Numerical trajectories demonstrate that the internal entropy production can be considered as a Lyapunov function candidate to assess the stability for the nonlinear irreversible flash-drum model. 

\subsection{Change of Coordinates}
\label{sec:diff-algebr-syst}

Extensive variables in \eqref{eq:differential} are considerably larger in magnitude than the intensive variables in the interface description \eqref{eq:algebraic}.  Such differences are known to cause the Jacobian matrix of the system to be ill-conditioned \citep{RITSCHEL2018281}. This leads to precision problems during the numerical integration of a DAE or an ODE system. To avoid scale differences between the bulk-phase and the interface models, we use standard definitions for molar fractions and molar concentration together with constitutive equations \eqref{eq:volume}-\eqref{eq:internalEnergy} to write a change of coordinates
  \begin{multline}
    (N_{g,1} \ldots N_{g,c} \;\; N_{l,j} \ldots N_{l,c} \;\; U_g \;\;
    U_l \;\; K_g \;\; K_l) \\\mapsto (y_1 \ldots y_{c-1} \;\; x_1
    \ldots x_{c-1} \;\; T_g \;\; T_l \;\; F_{g,V,\mathrm{out}} \;\; F_{l,V,\mathrm{out}}
    \;\; C_g \;\; V_l), \nonumber
  \end{multline}
given by
  \begin{subequations}\label{eq:intensive_variables}
    \begin{align}
      y_j &=  N_{g,j} / N_g  &j\in\{1,\ldots,c-1\}\\
      x_j &=  N_{l,j} / N_l  &j\in\{1,\ldots,c-1\}\\
      T_\alpha &= T_{\text o} + (U_\alpha - U_{\alpha,o}) / \mathcal C_\alpha &\alpha \in \{g,l\} \label{eq:temperature}\\
      F_{\alpha, V, \mathrm{out}} &= A_{\alpha,\mathrm{out}} \sqrt{  2 K_\alpha / M_\alpha }, &\alpha \in \{g,l\}  \label{eq:volumetric_outflow}\\    
      C_g &= N_{g} / (V_o - V_l) \label{eq:concentration_gas}\\ 
      V_l &= \bar{v} N_l. \label{eq:volume_liquid_again}
    \end{align}
  \end{subequations}
Here $N_\alpha$ and $M_\alpha$ represent the total molar and mass holdups for phase $\alpha$, $U_{\alpha,o}$ is a reference state at temperature $T_o$, $\mathcal C_\alpha$ is the heat capacity for phase $\alpha$, $V_o$ is the volume of the flash-drum, and $\bar{v} = \sum \bar{\text v}_j x_{j}$ represents the liquid molar volume.

As the change of coordinates is bijective (\ref{sec:jacobian-change-coordinates}), time differentiation of~\eqref{eq:intensive_variables} leads to a description of the flash-drum equivalent to \eqref{eq:DAE} with the advantage that the bulk-phase and the interface variables have the same order of magnitude.  Taking a time derivative on both sides of~\eqref{eq:intensive_variables} allows us to rewrite the DAE~\eqref{eq:DAE} as%
  \begin{subequations}\label{eq:DAE_again}
    \begin{align}
      \dot {\boldsymbol z} &= \text M^{-1}\; h({\boldsymbol z},{\bf w}) \label{eq:differential_again}\\
      0 &= g({\boldsymbol z},{\bf w}), 
    \end{align}
  \end{subequations}
where 
  \begin{align}
    {\boldsymbol z} = [y_1 \ldots y_{c-1} \;\, x_1 \ldots x_{c-1} \;\, T_g \;\, T_l \;\, F_{l,V,\mathrm{out}} \;\, F_{g,V,\mathrm{out}} \;\, C_g\;\, V_l]^{\text t} \nonumber
  \end{align}
represents the new bulk-phase variables, and %
  \begin{align}
    {\bf w} = [y_{1,i} \ldots y_{c,i} \;\; x_{1,i} \ldots x_{c,i} \;\; T_i \;\; n]^{\text t} \nonumber
  \end{align}
denotes again the interface variables. The matrix $\text M$ in \eqref{eq:DAE_again} is diagonal of dimension $2c+4$ 
  \begin{align}
    \text M = \mathrm{diag}\bigg[  C_g V_g 1_{c-1} \;\;\;  C_l V_l 1_{c-1} \;\;\; \mathcal C_{g} \;\;\; \mathcal C_{l} \;\;\; \frac{ M_g v_g }{A_{g,\mathrm{out}}} \;\;\; \frac{ M_l v_l }{A_{l,\mathrm{out}}} \;\;\; V_g  \;\;\; 1 \bigg]
  \end{align}
where $1_{c-1}$ stands for a row vector of dimension $c-1$ with ones as elements. The entries in the vector function $h$ in \eqref{eq:DAE_again} are given by
  \begin{align}
    h_j 
    &= F_{g,N,\mathrm{in}} \left( y_{j,\mathrm{in}} - y_j \right) + k_{g,j} C_g (y_{j,i} - y_j) &j\in\{1,\ldots,c-1\}\nonumber\\
    h_{j+c-1} 
    &= F_{l,N,\mathrm{in}} \left( x_{j,\mathrm{in}} - x_j \right) - k_{l,j} C_l (x_j - x_{j,i}) &j\in\{1,\ldots,c-1\}\nonumber\\
    h_{2c-1}  
    &= F_{g,N,\mathrm{in}} \bar{\mathcal C}_{g,\mathrm{in}} \big( T_{g,\mathrm{in}} - T_g \big)  + F_{g,V,\mathrm{in}} P_{g,\mathrm{in}} - F_{g,V,\mathrm{out}} P_g \nonumber\\
    &+ \lambda_{g}(T_{g,Q} - T_g) + \lambda_{g,i} (T_i - T_g) - P_g \dot V_g + P_g n_{g,V} \nonumber\\
    h_{2c} 
    &= F_{l,N,\mathrm{in}} \bar{\mathcal C}_{l,\mathrm{in}} \big( T_{l,\mathrm{in}} - T_l \big)  + F_{l,V,\mathrm{in}} P_{l,\mathrm{in}} - F_{l,V,\mathrm{out}} P_l \nonumber\\
    & + \lambda_{l}(T_{l,Q} - T_g) - \lambda_{l,i} (T_l - T_i) - P_l \dot V_l - P_l n_{l,V}  \nonumber\\
    h_{2c+1} 
    &= \big( 0.5(v_{g,\mathrm{in}}^2 - v_g^2)\rho_{g,\mathrm{in}} + P_{g,\mathrm{in}} \big) F_{g,V,\mathrm{in}}  \nonumber\\
    & - P_g (F_{g,V,\mathrm{out}} + \dot V_g) + \big( 0.5(v_{i,g}^2 - v_g^2) + P_g/\rho_g \big) m\nonumber\\
    h_{2c+2} 
    &= \big( 0.5(v_{l,\mathrm{in}}^2 - v_l^2)\rho_{l,\mathrm{in}} + P_{l,\mathrm{in}} \big) F_{l,V,\mathrm{in}}  \nonumber\\
    & - P_l (F_{l,V,\mathrm{out}} + \dot V_l)  - \big( 0.5(v_{i,l}^2 - v_l^2) + P_l/\rho_l \big) m\nonumber\\
    h_{2c+3} 
    &= F_{g,V,\mathrm{in}}C_{g,\mathrm{in}} - F_{g,V,\mathrm{out}} C_g + n - C_g \dot{V_g} \nonumber\\
    h_{2c+4} 
    &= F_{l,V,\mathrm{in}}C_{l,\mathrm{in}} \bar{\text v}_{\mathrm{in}} - F_{l,V,\mathrm{out}} C_l \bar{\text v} - n_V, \nonumber
  \end{align}
where sub-index ``$\mathrm{in}$'' refers to inflows, $\bar{\mathcal C}_{\alpha}$ represents molar heat capacity for the flow $\alpha$, $m = \sum \bar m_j n_j$ holds for the interface mass exchange rate, $n_{l,V} = \sum \bar{\text v}_j n_j$ represents the volumetric change rate at which the liquid phase evaporates, and $ P_g n_{g,V} = n\,R\,T_g $. The algebraic restriction $g$ in~\eqref{eq:DAE_again}, defined before in  \eqref{eq:DAE}, is rewritten here for the sake of clarity
  \begin{align}
    g_j   
    &= k_{g,j} \, C_g (y_{j,i} - y_j) -  k_{l,j} \, C_l (x_j - x_{j,i}) + (y_j -x_j)n & j \in \{ 1,\ldots,c-1 \} \nonumber\\
    g_{c} 
    &= \sum_{j=1}^c n_{j} \Delta \bar h_{\mathrm{vap}} + \lambda_{g,i} (T_{g} - T_{i}) - \lambda_{l,i} (T_{l} - T_{i}) \nonumber\\
    g_{j+c}
    &= y_{j,i} - K_j(T_i,x_{1,i},\ldots,x_{c,i}) x_{j,i} &j \in \{ 1,\ldots,c \} \nonumber\\
    g_{2c + 1} 
    &= 1 - \textstyle\sum_{j=1}^c x_{j,i} \nonumber\\
    g_{2c+2} 
    &= 1 - \textstyle\sum_{j=1}^c y_{j,i}. \nonumber
  \end{align}
Equation \eqref{eq:DAE_again} represents again an index one DAE system describing the irreversible flash-drum.  The system has $2c + 4$ bulk-phase variables and $2c+2$ interface variables.  Conservation principles applied to each bulk-phase give  $2c+4$ differential equations, while the interface is described by $2c+2$ algebraic equations.  In contrast with \eqref{eq:DAE}, all the variables in \eqref{eq:DAE_again} have the same order of magnitude, reducing the possibility of numerical integration problems.

\subsection{Local Stability}

In order to study the dynamic properties of the proposed model, we briefly review local stability results for linear index one DAE systems. A complete view on stability for DAE systems can be found in the work by \citet{yangTCS2013}.  

Let $({\boldsymbol z}^\star,{\bf w}^\star)$ be an equilibrium point of \eqref{eq:DAE}, \textit{i.e.},%
  \begin{align}
    0 &= h({\boldsymbol z}^\star,{\bf w}^\star) \nonumber\\
    0 &= g({\boldsymbol z}^\star,{\bf w}^\star). \nonumber
  \end{align}
As the Jacobian $J_{\bf w}(g)$ is full ranked (see \ref{sec:jacobian-g}), we can write a linearized version of \eqref{eq:DAE_again} around as $({\boldsymbol z}^\star,{\bf w}^\star)$ as
  \begin{align}
    \dot{\boldsymbol z}
    &= \Theta \; (\boldsymbol{z} - \boldsymbol{z}^\star). \label{eq:linearized_DAE}
  \end{align}
where the matrix $\Theta$ corresponds to the linearization of the right hand side terms in \eqref{eq:differential_again} at the equilibrium point
  \begin{align}
    \text M_l = \left[ J_{\boldsymbol z}(\text M^{-1} h) - J_{\bf w}( \text M^{-1} h) \; [J_{\bf w}(g)]^{-1} \; J_{\boldsymbol z}(g) \right] \big|_{({\boldsymbol z}^\star,{\bf w}^\star)}. \label{eq:M_lh}
  \end{align}
Lyapunov's first method states that if the spectrum of the matrix (\ref{eq:M_lh}) is contained in the left half-complex plane, then \eqref{eq:DAE_again} is locally asymptotically stable in a neighborhood of the stationary state  $({\boldsymbol z}^\star,{\bf w}^\star)$. 

\subsection{Case Study}
\label{sec:case-study-1}

A non-ideal methanol-water mixture is considered to illustrate the proposed model and its analysis. With two components, the DAE system (\ref{eq:DAE_again}) is given by 14 equations and 14 variables. Fixing the inflows to be at thermodynamic equilibrium $(T^\star,\, P^\star,\, y^\star_1,\, y^\star_2,\, x^\star_1,\, x^\star_2)$, the stationary state for (\ref{eq:DAE_again}) corresponds to
  \begin{align}
    {\boldsymbol z}^\star &= ( y_1^\star,\, x_1^\star,\, T^\star,\, T^\star,\, F_{g,V}^\star,\, F_{l,V}^\star,\, C^\star_g,\, V^\star_l ) \nonumber \\
    {\bf w}^\star &= ( y_{1}^\star,\, y_{2}^\star,\, x_{1}^\star,\, x_{2}^\star,\, T^\star,\, 0 ), \nonumber
  \end{align}
where $C^\star_g$ is determined from Equation (\ref{eq:P_g}), $V^{\star}_l$ is the volume of the liquid phase at equilibrium, and $F_{\alpha,V}^\star$ represents the stationary state volumetric inflow rates.  To determine numerical values for the steady state, inflow properties are assumed to be at thermodynamic equilibrium at $T^\star = 351.24 \mathrm{K}$ (78.09$^o\mathrm{C}$) and $P^\star =  101.3 \mathrm{kPa}$ (0.9998 $\mathrm{atm}$). Liquid-vapor equilibrium is cal\-culated using an Antoine--Margules thermodynamic model (see \ref{sec:example_margules}). In addition, inflows are fixed at $1\; \mathrm{m}^3/s$, and the liquid phase is set to occupy 10\% of the total volume $V_T = 1\; \mathrm{m}^3$. At these conditions, the stationary state $({\boldsymbol z}^\star,{\bf w}^\star)$ takes the values given in Table~\ref{tab:thermo_equilibrium}.
\begin{table}[h!]
  \centering
  \begin{tabular}{cccc}
    \hline
    $y_1^\star = 0.6615$, &$y_2^\star = 0.3385$ &$T^\star = 78.09 \;^o\mathrm{C}$, & $V_l^\star = 0.1\;\mathrm{m}^3$ \\
    $x_1^\star = 0.2764$, &$x_2^\star = 0.7236$ &$C_g^\star = 34.6874 \;\mathrm{mol}/\mathrm{m}^3$, & $F_{\alpha,V}^\star = 1\; \mathrm{m}^3/\mathrm{s}$\\
    \hline
  \end{tabular}
  \caption{Numerical values for the stationary state}
  \label{tab:thermo_equilibrium}
\end{table}

\subsubsection{Stability Analysis}\label{sec:stability-analysis}

Below we assess local stability when the system \eqref{eq:DAE_again} is linearized through \eqref{eq:linearized_DAE} at the stationary state given in Table \ref{tab:thermo_equilibrium}.  

\begin{enumerate}[label=S\arabic*]
\item {Scenario 1  (non-isobaric operation regime).} 
At equilibrium, the linearized system \eqref{eq:linearized_DAE} has rank 8. This system is unstable as the spectrum of the system has two positive eigenvalues $\lambda_{1}\approx 7.7691 \times 10^{-3}$, and $\lambda_{2}\approx 1 \times 10^{-12}$. The remaining eigenvalues are contained in the left half plane, between $\lambda_{3} \approx -1$ and $\lambda_{8} \approx -1.2 \times 10^{5}$.

\item {Scenario 2 (isobaric operation regime)}.  We calculate the stability properties of the system after a perfect pressure controller is included in \eqref{eq:DAE_again} making pressure constant at $P_l = P_g = P^\star$.  It must  be noted here that the introduction of the perfect pressure controller changes the dynamic description. Gas concentration trajectories are described under an isobaric regime by the algebraic restriction
  \begin{align}
    0 
    & = \frac{1}{R} \dfrac{dP_g}{dt} =  C_g \dfrac{dT_g}{dt} + T_g \dfrac{dC_g}{dt} = \dfrac{T_g}{\mathcal C_g} h_{2c-1} +  \dfrac{C_g}{V_g} h_{2c+3}, \label{eq:isobaricRestriction}
  \end{align}
where functions $h_{2c-1}$ and $h_{2c+3}$ represent the right hand terms for the temperature and concentration derivatives in \eqref{eq:differential_again}:
  \begin{align}
    h_{2c-1}  
    &:= F_{g,N,\mathrm{in}} \bar{\mathcal C}_{g,\mathrm{in}} \big( T_{g,\mathrm{in}} - T_g \big)  + F_{g,V,\mathrm{in}} P_{g,\mathrm{in}} - F_{g,V,\mathrm{out}} P_g \nonumber\\
    &+ \lambda_{g}(T_{g,Q} - T_g) + \lambda_{g,i} (T_i - T_g) - P_g \dot V_g + P_g n_{g,V} \nonumber\\    
    h_{2c+3} 
    &:= F_{g,V,\mathrm{in}}C_{g,\mathrm{in}} - F_{g,V,\mathrm{out}} C_g + n - C_g \dot{V_g}. \nonumber
  \end{align}
Equation \eqref{eq:isobaricRestriction} can be solved to write $C_g$ as a state function. Then, the dimension of 
the linearized flash-drum is reduced as $C_g$ is an equation of state for the isobaric system and not a state variable as in the non-isobaric model \eqref{eq:DAE_again}.  Under the isobaric restriction \eqref{eq:isobaricRestriction} the linearized system \eqref{eq:linearized_DAE} has rank 7.  

One eigenvalue for \eqref{eq:linearized_DAE} restricted by the isobaric equation \eqref{eq:isobaricRestriction} seems to be positive,  $\lambda_1 \approx 1 \times 10^{-13}$.  The rest of the spectrum is contained in the negative half-line between $\lambda_{2} \approx -1$ and $\lambda_{7} \approx -1.2 \times 10^{5}$. The positive eigenvalue is so close to zero that we cannot draw a sound conclusion regarding stability as numerical error can be the underlying cause for the positivity of $\lambda_1$.  
\end{enumerate}

\subsubsection{Numerical Simulations for Isobaric Operation}

To further investigate the stability of the isobaric regime, three dynamic simulations for Scenario 2 are now discussed.  For the first scenario, the system is initialized at stationary state and the inflow liquid temperature is reduced, pushing the system far from the thermodynamic equilibrium state.  In the second simulation, the system is disturbed from the stationary state. The disturbance is then removed at time $t=2\mathrm{s}$ and the system goes back to the equilibrium state.  For the last simulation we show how the internal entropy decreases as the system goes back to thermodynamic equilibrium starting from non-equilibrium initial states.

\begin{figure}[h!]
  \centering
  \includegraphics[width=\linewidth]{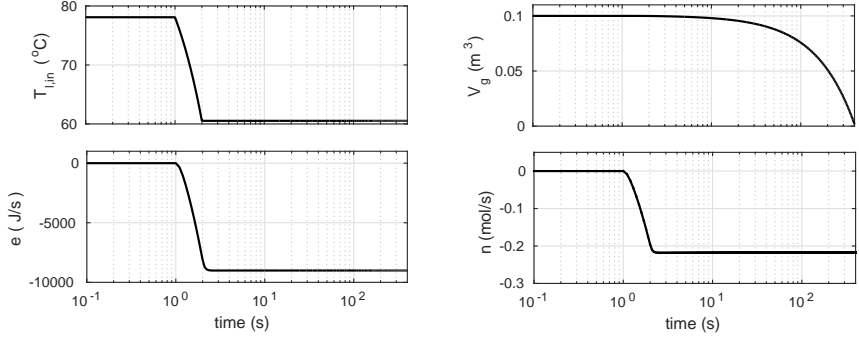}
  \caption{Disturbance, liquid volume and interface flow rates for numerical simulation 1. }
  \label{fig:depleted}
\end{figure}

\emph{Numerical simulation 1 (Figure~\ref{fig:depleted}).} The inflow liquid temperature is disturbed through a ramp disturbance for $1\leq t < 2$. The liquid inflow reaches a temperature $T_{l,\mathrm{in}} = 0.95 \times T^\star$ for $t\ge 2$ and the liquid-vapor system is forced to operate far from thermodynamic equilibrium.  As the system remains far from thermodynamic equilibrium,  mass and energy flow from the gas to the liquid phase.  The gas phase condensates completely around $t\approx 460 \mathrm{s}$.  For better appreciation of the dynamic behavior Figure~\ref{fig:depleted} is presented using a log scale for time.

\begin{figure}[h!]
  \centering
  \includegraphics[width=\linewidth]{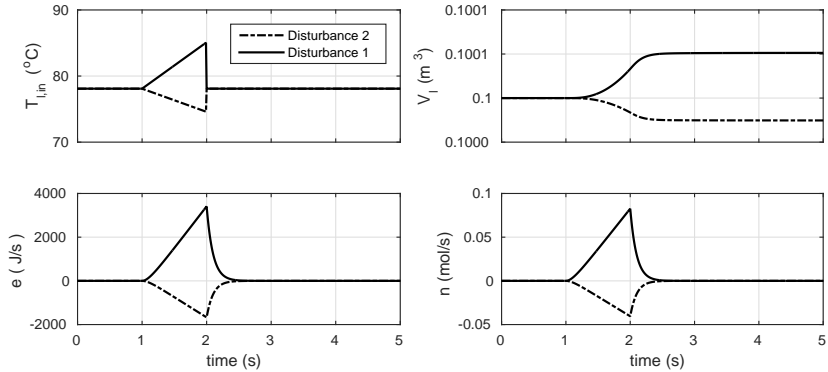}
  \caption{Disturbance, liquid volume and interface flow rates for numerical simulation 2. }
  \label{fig:disturbance_interface_flowrates}
\end{figure}

\begin{figure}[h!]
  \centering
  \includegraphics[width=\linewidth]{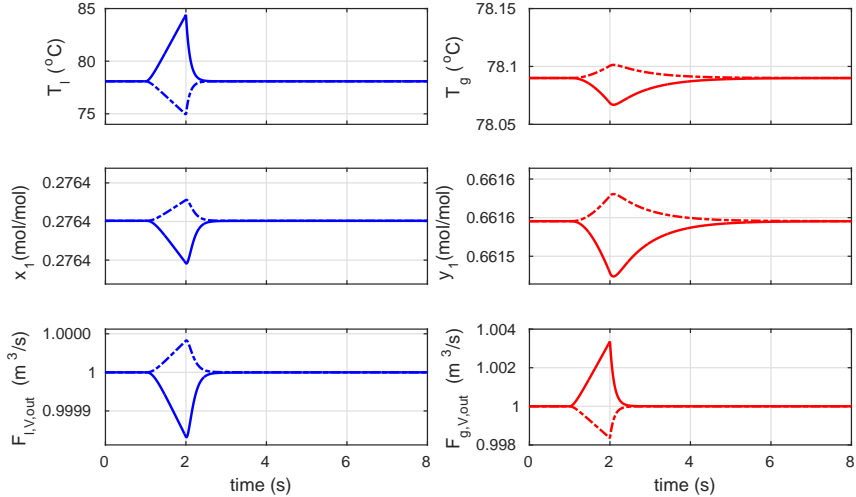}
  \caption{Dynamic response for the methanol-water irreversible flash-drum against ramp disturbances: temperature, molar composition (methanol) and volumetric outflow trajectories against time; red plots (right) represent the gas phase, and blue plots (left) represent the liquid phase}
  \label{fig:TXYFV_response}
\end{figure}

\emph{Numerical simulation 2 (Figure~\ref{fig:disturbance_interface_flowrates} and Figure~\ref{fig:TXYFV_response}).} A ramp disturbance is introduced in the liquid inflow temperature for $1\leq t < 2$.  In contrast with the numerical simulation 1, the disturbance is removed at $t=2 \mathrm{s}$ and the inflow temperature is set back its nominal value $T_{l,\mathrm{in}} = T^\star$, see Figure~\ref{fig:disturbance_interface_flowrates}.  This scenario is studied through two different disturbances.  It can be seen in Figure~\ref{fig:TXYFV_response} that inhomogeneities in temperature and composition appear between phases as a consequence of the disturbance.  Then, transfer processes redistribute the mass and the energy in the system as the flash-drum goes back to the equilibrium state.

\begin{figure}[h!]
  \centering
  \includegraphics[width=0.75\linewidth]{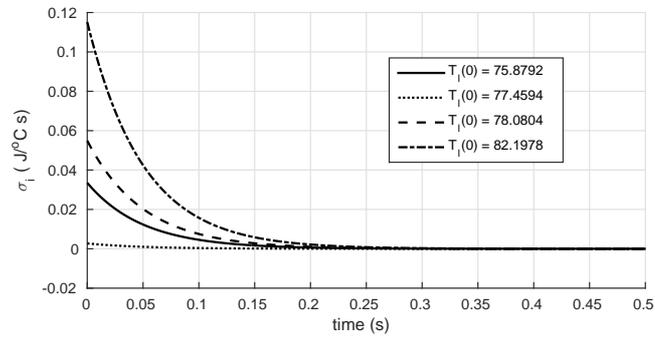}
  \caption{Internal entropy production rate for the isobaric methanol-water irreversible flash-drum starting far from thermodynamic equilibrium $(T_{eq} = 78.09^\circ C, P_{eq} = 101.3 kPa)$.}
  \label{fig:entropy_production}
\end{figure} 

\emph{Numerical simulation 3 (Figure~\ref{fig:entropy_production}).} 
Results from the Numerical simulation 2 point towards the equilibrium state being a stable steady state (Figure~\ref{fig:TXYFV_response}), despite the presence of a zero eigenvalue in the linearized system (see Section~\ref{sec:stability-analysis}).  To extend the analysis, the last simulation here presented tests the internal entropy production \eqref{eq:sigma_i} as a Lyapunov function candidate for the irreversible flash-drum\footnote{Dynamic trajectories on the internal entropy production \eqref{eq:sigma_i} are computed using the ideal gas model and the fugacity to calculate the chemical potentials (\ref{sec:chemical-potential}).}.  In this scenario, the liquid-vapor system starts from an initial condition far from thermodynamic equilibrium while the inflows remain constant at the conditions in Table~\ref{tab:thermo_equilibrium}.  It can be seen that the internal entropy production, Equation~\eqref{eq:sigma_i}, behaves as a Lyapunov function as the irreversible flash-drum goes back to the stationary state, see Figure \ref{fig:entropy_production}.  This suggests again that the thermodynamic equilibrium state is a stable stationary state for the irreversible flash-drum.  

\section{Conclusions and Future Work}
\label{sec:concl-future-work}

In this article, modeling aspects of dynamic flash-drum systems are explored using a non-equilibrium physics-based model.  The description here presented considers transport phenomena as well as conservation principles to write the dynamics of a multiphase system as a nonlinear DAE system of index one.  The proposed dynamic model describes the evolution of liquid and vapor phases as separated sub-systems interconnected through an interface.  The introduction of the interface exchange rates in the model can even predict the collapse of one phase for systems that operate consistently far from equilibrium.  Moreover, the model presented here describes how entropy is produced as a consequence of external (mass/energy exchanges between the system and the environment) and internal phenomena (mass/energy exchanges between phases).  

Numerical evidence shows that the linearized irreversible flash-drum system has positive eigenvalues for a non-isobaric operation regime.  For the isobaric case, trajectories appear stable in simulations as the internal entropy production appears to be a Lyapunov function candidate for the system.  The numerical results pinpoint the need to perform a deeper analysis regarding the stability for the nonlinear liquid-vapor irreversible system.  In future research, a nonlinear passivity-based stability analysis approach for the DAE model along the lines of \citet{garcia_sandoval2015} will be considered to get an input-output perspective on the analysis and control problem for multiphase chemical systems.

\section*{Acknowledgments}
The research presented in this paper is supported by the Mexican Council for Science and Technology (Grant 410828), and the Institute of Information and Communication Technologies, Electronics and Applied Mathematics (ICTEAM) at Universit\'e catholique de Louvain, Belgium.


\bibliographystyle{elsarticle-num-names}
\bibliography{references}             

\newpage
\appendix
\section{Jacobians}\label{sec:jacobians}

\subsection{Jacobian for the algebraic system g}\label{sec:jacobian-g}

\noindent The Jacobian of $g$, see Equation \eqref{eq:DAE}, with respect to 
  \begin{align}
    {\bf w} = [y_{1,i} \ldots y_{c,i} \;\; x_{1,i} \ldots x_{c,i} \;\; T_i \;\; n]^{\text t} \nonumber
  \end{align}
corresponds to the full rank\footnote{The Jacobian matrices and the respective ranks in \ref{sec:jacobians} are calculated using the open source computer algebra system Maxima version:5.32.1.} $2c+2$ square matrix
  \begin{align}
    \dfrac{\partial g}{\partial {\bf w}} 
    &= 
      \left[
      \begin{array}{cccc}
        C_g  \text k_g & C_l \text k_l & 0_{(c-1, 1)} & {\text y}_{c-1} - {\text x}_{c-1} \\ \\
        C_g \Delta \text h_g \text k_g & C_l \Delta \text h_l \text k_l &\lambda_{g,i} + \lambda_{l,i}  & {\text y}_{c}^{\text t} \text h_g - {\text x}_{c}^{\text t} \text h_l \\ \\
        I - {\text x_i} \, J_{\bf y_i} (K) & - \text K - {\text x_i} \, J_{\bf x_i} (K) & - {\text x_i} \, J_{T_i} (K) & - {\text x_i} \, J_{n} (K) \\ \\
        0_{(1,c)} & 1_{(1,c)} & 0 & 0 \\ \\
        1_{(1,c)} & 0_{(1,c)} & 0 & 0 \label{eq:Jg}
      \end{array} 
      \right].
  \end{align}
The term $\text k_\alpha$ in \eqref{eq:Jg} represents a $(c-1) \times c$ diagonal sub-matrix with an additional column of zeros 
  \begin{align}
    \text k_\alpha = 
    \begin{bmatrix}
      k_{\alpha,1} & \cdots & 0 & 0 \\
      \vdots &\ddots &\vdots &\vdots \\
      0 & \cdots & k_{\alpha,c-1} & 0
    \end{bmatrix}.\nonumber
  \end{align}
The terms ${\text y}_c$ and ${\text x}_c$ stand for a column vectors
  \begin{align}
    {\text y}_c = [y_1 \; \ldots \; y_c]^{\text t}, \phantom{....}
    {\text x}_c = [x_1 \; \ldots \; x_c]^{\text t}. \nonumber
  \end{align}
When subindex $c-1$ is used to write ${\text y}_{c-1}$, and ${\text x}_{c-1}$, the composition vector contains only the first $c-1$ molar fractions. The term $\Delta \text h_\alpha$ represents a row vector
  \begin{align}
    \Delta \text h_\alpha = [\bar h_{\alpha,1}-\bar h_{\alpha,c}  \; \ldots \; \bar h_{\alpha,c-1}-\bar h_{\alpha,c}],
  \end{align}
and the vector $\text h_\alpha$ contains the enthalpies for the components in the mixture
  \begin{align}
    {\text h}_\alpha = [\bar h_{\alpha,1} \; \ldots \; \bar h_{\alpha,c}]^{\text t}.
  \end{align}
The symbol $\text x_i$ stands for a diagonal matrix with interface molar compositions
  \begin{align}
    \text x_i =  \mathrm{diag}[x_{1,i} \; \ldots \; x_{c,i}],
  \end{align}
and $J_{\beta} (K)$ holds for the Jacobians of the equilibrium ratio $K$ with respect to $\beta \in \{y_{1,i} \ldots y_{c,i} \;\; x_{1,i} \ldots x_{c,i} \;\; T_i \;\; n \}$. Zeros and ones in \eqref{eq:Jg} represent zero vectors, one vectors and scalars when dimension is not specified.

\subsection{Jacobian for the change of coordinates}\label{sec:jacobian-change-coordinates}

\noindent To show that that Equation \eqref{eq:intensive_variables} is bijective, it is enough to demonstrate that the Jacobian matrix of the mapping is non-singular. To do so, we rewrite the change of coordinates as 
  \begin{multline}
    F : (N_{g,1} \ldots N_{g,c} \;\; N_{l,j} \ldots N_{l,c} \;\; U_g \;\;
    U_l \;\; K_g \;\; K_l) \\\mapsto (y_1 \ldots y_{c-1} \;\; x_1
    \ldots x_{c-1} \;\; T_g \;\; T_l \;\; F_{g,V,\mathrm{out}} \;\; F_{l,V,\mathrm{out}}
    \;\; C_g \;\; V_l), \nonumber     
  \end{multline}
where
  \begin{subequations}
    \begin{align}
      F_j &=  N_{g,j} / N_g,  &j\in\{1,\ldots,c-1\} \nonumber\\
      F_{j+c-1} &=  N_{l,j} / N_l, &j\in\{1,\ldots,c-1\} \nonumber\\
      F_{2c-1} &= T_{\text o} + (U_g - U_{g,o}) / \mathcal C_g \nonumber\\
      F_{2c} &= T_{\text o} + (U_l - U_{l,o}) / \mathcal C_l \nonumber\\
      F_{2c+1} &= A_{g,\mathrm{out}} \sqrt{  2 K_g / M_g }, \nonumber\\    
      F_{2c+2} &= A_{l,\mathrm{out}} \sqrt{  2 K_l / M_l },\nonumber \\    
      F_{2c+3} &= N_{g} / (V_o - V_l) \nonumber\\ 
      F_{2c+4} &= \bar{\text v} N_l. \nonumber
    \end{align}
  \end{subequations}
Then, the Jacobian of $F$ with respect to 
  \begin{align}
    {\bf z} = [N_{g,1} \ldots N_{g,c} \;\; N_{l,j} \ldots N_{l,c} \;\; U_g \;\; U_l \;\; K_g \;\; K_l]\nonumber
  \end{align}
corresponds to the full rank sparse matrix
  \begin{align}
    &\dfrac{\partial F}{\partial {\bf z}} 
    =
    \begin{bmatrix}
      \dfrac{1}{N_g} \Theta_g(y) &0_{(c-1,c)} &0_{(c-1,1)} &0_{(c-1,1)}  &0_{(c-1,1)} &0_{(c-1,1)} \\ \\
      0_{(c-1,c)} &\dfrac{1}{N_l} \Theta_l(x) &0_{(c-1,1)} &0_{(c-1,1)}  &0_{(c-1,1)} &0_{(c-1,1)} \\ \\
      -\dfrac{1}{\mathcal C_g} \text u_g &0_{(1,c)} &\dfrac{1}{\mathcal C_g} &0 &0 &0  \\ \\
      0_{(1,c)} &-\dfrac{1}{\mathcal C_l} \text u_l&0 &\dfrac{1}{\mathcal C_l} &0 &0  \\ \\
      -\dfrac{F_{g,V,\mathrm{out}}}{M_g} \bar{\text m} &0_{(1,c)} &0 &0 &\dfrac{F_{g,V,\mathrm{out}}}{K_g} &0 \\ \\
      0_{(1,c)} &-\dfrac{F_{l,V,\mathrm{out}}}{M_l} \bar{\text m} &0 &0 &0 &\dfrac{F_{l,V,\mathrm{out}}}{K_l} \\ \\      
      0_{(1,c)} &\bar{\text v} &0 &0 &0 &0 \\ \\
      -\dfrac{1}{V_g} \cdot 1_{(1,c)} & -C_g \bar{\text v} &0 &0 &0 &0 \\ \\
    \end{bmatrix}, \label{eq:JF}
  \end{align} 
where, $\Theta_g(\cdot)$ stands for a $(c-1)\times c$ matrix
  \begin{align}
    \Theta_g = 
    \begin{bmatrix}
      1 - y_1 &-y_1 &\cdots &-y_1 &-y_1\\
      - y_2 & 1-y_2 &\cdots &-y_2 &-y_2\\
      \vdots &\vdots &\ddots &\vdots &\vdots\\
      -y_c &-y_c  &\cdots & 1 -y_c &-y_c\\
    \end{bmatrix} \nonumber    
  \end{align}
and $\Theta_l(\cdot)$ holds for an equivalent matrix in terms of liquid molar fractions. The molar energy, the liquid molar volume, and the molar mass vectors in \eqref{eq:JF} correspond to
  \begin{align}    
    \bar{\text u}_\alpha = [\bar u_{\alpha,1}, \ldots, \bar u_{\alpha,c} ], \;\;\;
    \bar{\text v} = [\bar{\text v}_{1}, \ldots, \bar{\text v}_{c} ], \;\;\;
    \bar{\text m} = [\bar{m}_{1}, \ldots, \bar{m}_{c} ]\nonumber
  \end{align}
Finally, the zeros and ones in \eqref{eq:JF} represent zero matrices/vectors, one vectors and scalars when dimension is not specified.

\section{Chemical potential and thermodynamic equilibrium}

\subsection{Chemical potential for gas and liquid mixtures}\label{sec:chemical-potential}

\noindent Consider a thermodynamic system formed by an ideal gas with internal energy $U$, and $N$ moles occupying a volume $V$. The entropy for that system satisfies \cite{C1985thermodynamics}
  \begin{align}
    \text S := N \bar s = N \bar{\text s}_{\text o} + N \mathtt c_{\text v} R \ln \bigg( \frac{U}{\mathtt c_{\text v} \, N R \, T_{\text o}} \bigg) + N R \ln \bigg( \frac{V}{N \bar{\text v}_{\text o}} \bigg), \nonumber
  \end{align}
where the reference state $\bar{\text s}_{\text o}$ corresponds to the molar entropy for a system at temperature $T_{\text o}$ with molar volume $\bar{\text v}_{\text o}$, and the dimensionless heat capacity $\mathtt c_{\text v} = c_{v}/R$ is considered a constant parameter. Values for $\mathtt c_{\text v}$ are reported in Table~\ref{tab:c_ideal_gas}.
\begin{table}[h!]
  \centering
  \begin{tabular}{ccc}\hline
    & Gas & Temperature range \\\hline
    $\mathtt c_{\text v} = 3/2$ &Monotonic non-interactive atoms &$T < 10^4$K \\
    $\mathtt c_{\text v} = 5/2$ &Diatomic non-interactive molecules &$T < 10^3$K \\
    $\mathtt c_{\text v} = 7/2$ &Diatomic non-interactive molecules &$T > 10^3$K \\\hline
  \end{tabular}  
  \caption{Dimensionless heat capacity $\mathtt c_{\text v}$ for ideal gases}
  \label{tab:c_ideal_gas}
\end{table}

For a mixture of ideal gases occupying a volume $\text V$ at temperature $T$, entropy is the sum of entropies that each component would have if it alone were to occupy the volume $\text V$ at temperature $T$ \cite[\S 3.4 -- Gibbs theorem]{C1985thermodynamics}. Then entropy for an ideal mixture of gases can be written as 
  \begin{align}
    &\text S = \sum_{j} N_j \bar{\text s}_{\text o, j} +  \mathcal C R \ln \bigg( \frac{U}{\mathcal C R T_{\text o}} \bigg) + N R \ln \bigg( \frac{V}{N \bar{\text v}_{\text o}} \bigg) - R \sum_{j} N_j \ln y_j, \label{eq:entropy_fundamental_mix_ideal_gases_fundanental}
  \end{align}
where the sum is taken over all the components in the mixture, $N = \sum N_j$ stands for the total molar holdup of the system, and $\mathcal C = \sum \mathtt c_{\text v, j} N_j$ represents the total heat capacity of the system. Note that Equation \eqref{eq:entropy_fundamental_mix_ideal_gases_fundanental} is the entropy fundamental equation defined by the thermodynamics formal structure \eqref{eq:formal_structure} for an ideal gas mixture
  \begin{align}
    \text S = S(U,V,N_1,\ldots,N_c). \nonumber
  \end{align}
It follows that chemical potential for component $j$ inside an ideal gas mixture can be calculated as the derivative of \eqref{eq:entropy_fundamental_mix_ideal_gases_fundanental}  with respect to mole numbers
\begin{subequations}\label{eq:entropy_fundamental_mix_ideal_gases}
    \begin{align}     
      &\dfrac{-\mu_j}{T} := \frac{\partial S}{\partial N_j} = \mu_j^\star(T,P) + RT \ln y_j\label{eq:entropy_fundamental_mix_ideal_gases_a}
    \end{align}
  where $\mu_j^\star(\cdot)$ represents the chemical potential of the ideal gas $j$, 
    \begin{align}
      &\mu_j^\star = -T \text s_{\text o, j} + R T ( \mathtt c_{\text v,j} + 1) -  R T \ln \bigg( \frac{P_{\text o}}{P} \bigg) \bigg( \frac{T}{T_{\text o}} \bigg)^{ \mathtt c_{\text v,j}+1}.
    \end{align}
\end{subequations}
Despite Equation \eqref{eq:entropy_fundamental_mix_ideal_gases_a} being only valid for ideal gas mixtures, it is a common practice to try to preserve this form  as far as possible when describing non-ideal systems \cite{prigogine1968introduction}. 

The chemical potential for a component $j$ inside a non-ideal mixture can be written as \cite{prigogine1968introduction}
  \begin{align}
    \mu_j =  \mu^{\mathrm{im}}_j(P,T) + RT\ln{\big(\gamma_j \big)},\nonumber
  \end{align}
where $\mu^{\mathrm{im}}$ represents the chemical potential of component $j$ inside an ideal mixture.  To measure deviations from ideal behavior, excess in chemical potential (referred by some authors as excess on partial
molar Gibbs potential $\bar g_i^{\mathrm{ex}}$) is defined as 
  \begin{align}
    \mu_j^{\mathrm{ex}} := \mu_j - \mu_j^{\mathrm{im}},\nonumber
  \end{align}
and thus the activity coefficient $\gamma_j$ is satisfies
  \begin{align}
    \mu_j^{\mathrm{ex}} = RT \ln \gamma_j. \nonumber
  \end{align}
Setting the chemical potential for $j$ inside the ideal system $\mu_j^{\mathrm{im}}$ to be represented by \eqref{eq:entropy_fundamental_mix_ideal_gases} we can write the chemical potential for component $j$ inside a liquid mixture as 
  \begin{align}
    \mu_j = \mu_j^\star(T,P) + RT \ln x_j + \mu_j^{\mathrm{ex}}.
  \end{align}
This description for the chemical potential has a clear physical interpretation.  The chemical potential $\mu_j$ is the chemical potential of $j$ as an ideal system (an ideal gas in this case), plus mixing effects (second term), plus a correction term that represents deviations from ideal behavior $\mu^{\mathrm{ex}}$.  In the following section, we write the excess in chemical potential $\mu_j^{\mathrm{ex}}$ using a liquid-vapor equilibrium model.

\subsection{Margules-Antoine equilibrium model} \label{sec:example_margules}

\noindent A liquid-vapor system with two components at temperature $T$ and pressure $P$ is said to be at thermodynamic equilibrium when the chemical potentials in liquid an gas phases are equal
  \begin{align}    
    \mu_{l,j} = \mu_{g,j} , \;\;\;\;\;\;\;\;\;\;\;\; &j \in\{1,2\}. \label{eq:equilibrium_chem_potential}
  \end{align}
As chemical potential is not measurable, it is common practice to rewrite \eqref{eq:equilibrium_chem_potential} as an equivalent equality between fugacities
  \begin{align}
    \bar f_j^{\,g} = x_{j} \, \gamma_j \, \bar f_j^{\,l}, \;\;\;\;\;\;\;\;\;\;\;\; &j \in\{1,2\}, \label{eq:equilibrium_fugacity}
  \end{align}
where $ \bar f_j^{\,g} = y_j P \label{eq:gas_fugacity}$, provided that $P$ is close to atmospheric pressure, stands as the gas fugacity for component $j$  in the gas mixture, and 
  \begin{align}
    x_j \gamma_j \bar f_j^{\,l} = x_j \, \exp \bigg( \frac{\mu_j^{\mathrm{ex}}}{RT} \bigg) \, \bigg( A_j - \frac{B_j}{T - C_j} \bigg). \nonumber
  \end{align}
Parameters $A_j,\,B_j,$ and $C_j$, in the previous equation stand as constants for Antoine's Equation.  The excess in chemical potential can be written as a function of a polynomial in the liquid composition $Q(x)$ \citep{T1993masstransfer}
  \begin{align}
    \mu_j^{\mathrm{ex}} = RT \bigg( -2Q + \dfrac{\partial Q}{\partial x_j} \bigg). \nonumber
  \end{align}
Given thermodynamic parameters $A_{12}$ and $A_{21}$, the liquid composition polynomial $Q$ can be written using Margules thermodynamic model as
  \begin{align}
    Q = x_1 x_2 ( A_{12} x_1 + A_{21} x_2). \nonumber
  \end{align}
Algebraic rearrangement of equation \eqref{eq:equilibrium_fugacity} leads to the nonlinear liquid-vapor equilibrium equation   
  \begin{eqnarray}
    \begin{bmatrix}
      y_{1} \\ y_{2}
    \end{bmatrix}
    =
    \begin{bmatrix}
      K_1 & 0 \\
      0 & K_2
    \end{bmatrix}
          \begin{bmatrix}
            x_{1} \\ x_{2}
          \end{bmatrix},   \label{eq:twoComponentEquilibrium}   
  \end{eqnarray}
where
  \begin{align}
    &K_1 = \frac1{P} 
      \left( A_1 - \frac{B_1}{T - C_1} \right)
      \exp \left( 2 A_{12} x_{1} x_{2} ( 1 - x_{1} ) + A_{21} x_{2}^2 ( 1 - 2x_{1} ) \right) \nonumber\\
    &K_2 = \frac1{P} 
      \left( A_2 - \frac{B_2}{T - C_2} \right)
      \exp \left( 2 A_{21} x_{1}  x_{2} ( 1 - x_{2}) + A_{12} x_{1}^2 ( 1 - 2 x_{2} ) \right). \nonumber
  \end{align}

\end{document}